\documentclass{INTERSPEECH2023}

\interspeechcameraready

\usepackage{amsmath,graphicx}

\usepackage{booktabs}
\usepackage{multirow}
\usepackage{subcaption}
\usepackage[belowskip=0pt,aboveskip=0pt]{caption}
\title{Joint speech and overlap detection: a benchmark over multiple audio setup and speech domains}
\name{Martin Lebourdais$^{1\star}$, Théo Mariotte$^{1,2\star}$,  Marie Tahon$^{1}$, Anthony Larcher$^{1}$, \\
\textit{Antoine Laurent$^{1}$, Silvio Montrésor$^{2}$,  Sylvain Meignier$^{1}$, Jean-Hugh Thomas$^{2}$}
\thanks{$^{\star}$ Both authors contributed equally.}
}
\address{$^{1}$LIUM, $^{2}$LAUM UMR 6613 IA-GS, Le Mans Université 
}
\email{\{first name\}.\{last name\}@univ-lemans.fr}

\begin{document}

\maketitle
 
\begin{abstract}
Voice activity and overlapped speech detection (respectively VAD and OSD) are key pre-processing tasks for speaker diarization.
The final segmentation performance highly relies on the robustness of these sub-tasks.
Recent studies have shown VAD and OSD can be trained jointly using a multi-class classification model.
However, these works are often restricted to a specific speech domain, lacking information about the generalization capacities of the systems.
This paper proposes a complete and new benchmark of different VAD and OSD models, on multiple audio setups (single/multi-channel) and speech domains (e.g. media, meeting...). 
Our 2/3-class systems, which combine a Temporal Convolutional Network with speech representations adapted to the setup, outperform state-of-the-art results.
We show that the joint training of these two tasks offers similar performances in terms of F1-score to two dedicated VAD and OSD systems while reducing the training cost. This unique architecture can also be used for single and multi-channel speech processing.
\end{abstract}

\section{Introduction}
Speaker diarization answers the question \textit{Who spoke and when?} in an audio stream. 
Today, this task remains difficult as shown by the numerous challenges recently organized \cite{ryant19_interspeech,yu2022m2met_short,iberspeech}.

Given an audio stream, speaker diarization pipelines generally address speech segmentation and speaker clustering in two distinct stages~\cite{ryant19_interspeech}.

Therefore, robust speech segmentation - mainly Voice Activity Detection (VAD) and Overlapped Speech Detection (OSD) - is essential to improve speaker diarization performance as shown in previous studies~\cite{garcia_perera_speaker_2020_short,bullock_overlap-aware_2020}.
VAD consists in segmenting an audio signal into speech and non-speech segments.

Several approaches have been proposed in the literature such as signal processing methods \cite{ghaemmaghami2010noise}, statistical models \cite{nemer2001robust}, and neural-based approaches \cite{lavechin_end--end_2020}.
OSD detects segments in which at least two speakers are simultaneously active.
Early studies mostly focus on statistical models \cite{boakye_improved_2011,yella2014overlapping} while recent approaches are mostly based on neural networks 
\cite{bullock_overlap-aware_2020,andrei2017detecting,lebourdais22_interspeech} and show promising results. 

While VAD and OSD have mainly been considered as two independent binary classification tasks,
they can be addressed jointly by considering three classes -- non-speech, single speaker, and overlapped speech -- according to the number of present speakers in each speech segment.
In \cite{jung21_interspeech}, such a 3-class problem is solved by training a recurrent convolutional network.
The use of far-field microphones and a Self-Attention Channel Combinator (SACC) feature extractor~\cite{mariotte22_interspeech} revealed the potential of spatial information for OSD.
\cite{Cornell2022} demonstrated that Temporal Convolutional Network (TCN) is well adapted for multiple-speaker activity detection with far-field microphones.



In this paper, we propose two 2-class VAD and OSD and 3-class VAD+OSD for mono and multi-channel signals.
We evaluate how beneficial is the 3-class approach in comparison to the use of two independent VAD and OSD models in terms of F1-score and training resources.
Each system is trained and evaluated on four different datasets covering various speech domains including both single and multiple microphone scenarios.
To the best of our knowledge, no benchmark has been conducted on these approaches across various speech domains and recording setups (multi/mono channel) in the literature.
%
%
%

This paper presents several contributions: It first claims a new state of the art for OSD on multiple corpora, it introduces a benchmark on 4 different datasets covering various speech domains in multi/mono channel scenarios and presents a reduction of the training cost using a 3-class approach with a detailed analysis of the benefits of this system.

\section{Datasets}
\label{sect:2_datasets}
Our benchmark datasets combine multiple speech domains including far-field audio recordings.
For each dataset, VAD and OSD labels are derived from the provided ground-truth segmentation. 
Table~\ref{tab:corpus_desc} summarizes corpus characteristics.

\begin{table}[ht!]
\centering
\caption{Corpus characteristics. $^\star$multi-microphone data.}
\begin{tabular}{@{}lccc@{}}
\toprule
Corpus   & Domain & Duration & Overlap prop. \\ \midrule
DIHARD   & Multiple    & 34~h      & 11.6\%             \\
ALLIES   & Media    & 328~h     & 3.2\%              \\
ALLIES-clean & Media & 6~h      & 13.9\%             \\
AMI$^\star$      & Meeting   & 100~h     & 24.7\%                   \\
CHiME-5$^\star$  & Dinner party   & 60~h      & 22.9\%            \\ \bottomrule
\end{tabular}

\label{tab:corpus_desc}
\end{table}

\vspace{-15pt}

\subsection{Single Channel}
Single channel experiments are conducted on 3 datasets: ALLIES~\cite{larcher:hal-03262914}, DIHARD~\cite{ryant19_interspeech} and AMI~\cite{Mccowan05theami_short}.
The ALLIES corpus is a soon-to-be-available French meta-corpus designed to gather and extend previous French data collected for diarization and transcription evaluation campaigns.
It consists of 328~h of audio extracted from 1998 to 2014 in 1008 shows with 5901 different speakers. 
The overlap proportion (in duration) fluctuates widely between broadcast news with little to no interaction and debates (around 10\% of overlaps). 
Despite a harmonization effort, the data collected and annotated under different protocols introduces some homogeneity problems~\cite{lebourdais:hal-03660323}. 
15 debate shows, referred to as ALLIES-clean, 
were selected in order to get a high overlap proportion, a manual and homogeneous speech segmentation, and diversity in the shows represented. 

The DIHARD corpus contains data from 7 domains with various recording qualities, situations, and degrees of spontaneity from read speech to phone conversations.
Since spontaneous speech naturally contains a high proportion of overlapped speech, this corpus is well-suited for OSD. 
This corpus is partitioned as intended for the challenge and evaluated on the official evaluation partition.

The AMI meeting corpus contains recordings of realistic meetings involving up to 5 participants in various environments.
The \textit{headset-mix} is used for single-channel experiments on this dataset. 
The data partition follows the protocol proposed in \cite{landini_but_2020}.

\subsection{Multiple Channels}
Multiple-channel experiments are conducted on 2 corpora: AMI~\cite{Mccowan05theami_short} and CHiME-5~\cite{barker2018fifth}.
We select AMI audio data captured by the \textit{Array 1} as a distant multi-microphone signal.
It consists of a uniform circular array (UCA) composed of 8 omnidirectional microphones placed in the center of the table during meetings.

The CHiME-5 dataset contains 20 dinner-party sessions involving 4 participants in a real-home environment. 
Speakers were asked to move between 3 rooms during the party.
Audio signals thus feature a strong background noise diversity with varying acoustic conditions.
Audio signals are captured with 6 linear arrays composed of 4 microphones.
For our experiments, only the first microphone of each array is selected.
Finally, the resulting signal contains 6~channels.

\vspace{-5pt}
\section{System overview}
\label{sect:3_system}
\begin{figure}[ht!]
    \centering
    \includegraphics[width=1\linewidth]{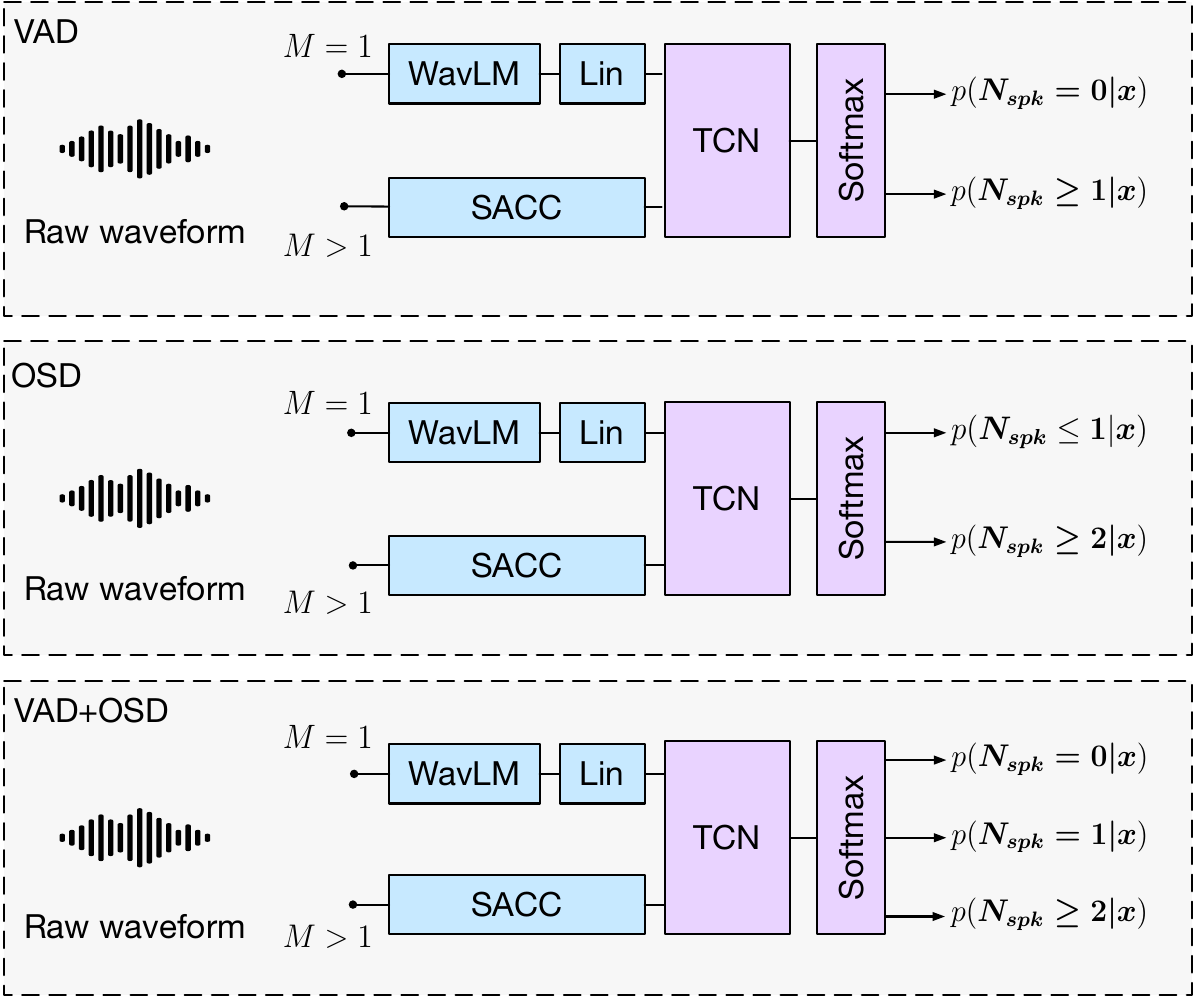}
    \caption{VAD, OSD, and VAD+OSD systems with the feature extractor  (blue) and the sequence modeling network (purple), $M$ is the number of channels.}
    \label{fig:archi}
\end{figure}

Figure~\ref{fig:archi} depicts an overview of VAD, OSD, and VAD+OSD systems.
While the feature extractor (in blue) is adapted with respect to the number of input channels, the sequence modeling network (in purple) processes the sequence of features before the frame classification.
The frame classification is done at a rate of 100~Hz, while the raw waveform is sampled at 16~kHz.
\vspace{-10pt}

\vspace{-10pt}

\subsection{Single channel features ($M=1$)}
The single channel feature extractor is based on the WavLM pre-trained model~\cite{chen_wavlm_2022_short}.
This choice is motivated by the performance obtained on the diarization task according to the SUPERB benchmark~\cite{yang_superb_2021_short}.
Furthermore, WavLM has been trained using simulated overlapped speech and is then more robust to this type of data. 
WavLM outputs speech representations every 20~ms. 
In order to align this representation with the target sequence, we decide to add a linear layer on top of the frozen WavLM.
The linear layer aims to transform a segment of 99 features extracted with WavLM over a 2~s window, into a 200-frame vector, supposedly aligned with our target.

\subsection{Multiple channel features ($M>1$)}

When multiple channels are available, feature extraction is performed using the Self-Attention Channel Combinator (SACC) \cite{gong_self-attention_2021_short}.
This architecture has previously shown its efficiency for OSD under distant speech conditions \cite{mariotte22_interspeech}.
The algorithm consists of a self-attention module \cite{vaswani_attention_2017_short} which computes per-channel weights from the multi-channel Short-Time Fourier Transform (STFT) of the input signal.
The channels are then weighted and combined in order to get a single-channel representation.
Combination weights are computed from the multichannel STFT calculated on 25~ms segments with 10~ms shift.
The attention module is composed of a single attention head of size $d=256$.
The combined representation is converted to the log-mel scale using $N_f=64$ filters.
Global Mean and Variance Normalization (MVN) is also applied before feeding the sequence modeling network.

\vspace{-5pt}

\begin{table*}[ht]
\centering
\caption{Overview of the F1-score (\%) for each system on the evaluation set of several corpora covering various domains, $^\star$ indicates multi-microphone data, $\dagger$ indicates that the results are taken from the original article.}
\setlength\tabcolsep{4.5pt}
\begin{tabular}{@{}rlcccccccccc@{}}

\toprule
            && \multicolumn{5}{c}{VAD} & \multicolumn{5}{c}{OSD} \\ 
            \cmidrule(l){3-7} \cmidrule(l){8-12}\\
            && DIHARD    & ALLIES    & AMI   & AMI$^{\star}$   & CHiME$^{\star}$    & DIHARD    & ALLIES   & AMI   & AMI$^{\star}$ & CHiME$^{\star}$\\ \midrule
\multirow{4}{*}{\rotatebox[origin=c]{90}{2-class}}&
VAD (ours)    & 97.0      &  99.8     & 97.4     & 96.4      & 99.8          & -         & -         & -         & -         & -\\
&OSD (ours)    & -         & -         & -        & -         & -          & \textbf{66.2}      & \textbf{71.6}  & \textbf{79.6}   & \textbf{72.2}      & \textbf{75.9}   \\
& Mel+CRNN~\cite{Cornell2022} & - & - & - & - & - & 51.3 & - & 66.0 & 57.2 & -\\ 
& Mel+TCN~\cite{cornell_detecting_2020} & - & - & - & - & - & 54.7 & - & 73.4 & 65.8 & -\\ \midrule
\multirow{4}{*}{\rotatebox[origin=c]{90}{3-class}}&
VAD+OSD (ours) & 97.0      & 89.2     &  97.2     & 96.6     &   99.3      & \textbf{66.8} & \textbf{75.4}      & \textbf{80.4}     & \textbf{71.8}      & \textbf{75.5} \\ 
& Mel+CRNN~\cite{Cornell2022} & - & - & - & - & - & 50.8 & - & 69.6 & 61.2 & -\\ 
& Mel+TCN~\cite{cornell_detecting_2020} & - & - & - & - & - & 54.5 & - & 73.8 & 67.9 & -\\ 
& SincNet+BLSTM~\cite{bredin:hal-03257524}$\dagger$ & - & - & - & - & - & 59.9 & - & 75.3 & - & -\\ \bottomrule
\end{tabular}
\label{tab:overview}
\end{table*}

\subsection{Sequence modeling and classification}

The sequence modeling network (in purple in Fig.~\ref{fig:archi}) takes as input a sequence $\boldsymbol{x}$ of single or multi-channel features and assigns a class to each frame of this sequence.
This task is performed using a TCN \cite{bai_empirical_2018} since this architecture has shown noticeable results on both VAD and OSD tasks \cite{lebourdais22_interspeech,mariotte22_interspeech,Cornell2022,cornell_detecting_2020}.
It is composed of 5 residual convolutional blocks repeated 3 times.
Classification is performed by a 1-d convolutional layer followed by a $\mathrm{softmax}$ activation function. 

For each frame in the output sequence, the VAD outputs the pseudo-probability to contain at least one speaker $p(N_{spk} \geq 1|\boldsymbol{x})$. 
The OSD outputs the pseudo-probability to contain speech from more than one speaker $p(N_{spk} \geq 2|\boldsymbol{x})$. 
Both VAD and OSD are then binary classifiers.
The joint VAD+OSD system outputs the pseudo-probability of either containing any speech $p(N_{spk}=0|\boldsymbol{x})$, speech from a single speaker $p(N_{spk}=1|\boldsymbol{x})$, or speech from more than one speaker $p(N_{spk}\geq2|\boldsymbol{x})$. 
The 3-class approach is then converted to 2-class VAD and OSD by merging the relevant classes.

\vspace{-5pt}
\subsection{Training and Evaluation}
In order to estimate the robustness over different speech domains, the three systems are trained and evaluated independently on the 5 datasets aforementioned. 
To counteract the small number of overlap segments, 50\% of the training segments are augmented on-the-fly by summing them to another randomly sampled training segment. Associated labels of each segment are also combined \cite{bredin_pyannoteaudio_2020_short}.
The loss function is a cross-entropy, and we used the ADAM optimizer with a learning rate of $lr=10^{-3}$.
Single-channel audio data is augmented with noise extracted from MUSAN~\cite{snyder2015musan} and additional reverberation using simulated room impulse responses.
Preliminary experiments have shown that data augmentation did not bring significant improvement in the far-field scenario. 


Following the DIHARD evaluation plan, we use the F1-score obtained on the evaluation set as a performance metric.
In the 2-class approach, only the positive class output ($N_{spk} \geq 1$ for VAD, and $N_{spk} \geq 2$ for OSD) is used for prediction and two detection thresholds are applied to predict binary labels \cite{bredin_pyannoteaudio_2020_short}.
In the 3-class approach, the class associated with the maximum $\mathrm{softmax}$ output is selected at the frame level. A working version of the code will soon be released\footnote{Hidden link for anonymous submission}. 

\section{Results}
\label{sect:4_results}
OSD and VAD results obtained on 5 single (DIHARD, ALLIES, AMI) and multi-channels (AMI$^\star$, CHIME-5$^\star$) datasets are presented in Table~\ref{tab:overview}.

\subsection{Single Channel}

So far, ALLIES corpus has only been studied for speaker diarization while discarding overlapping speech obtained in the manual reference~\cite{larcher:hal-03262914}. 
We provide the first evaluation of OSD for ALLIES data with a 71.6\% F1-score using the 2-class approach.

VAD performances are similar between the 2- and 3-class approaches except for the ALLIES data for which a strong F1-score degradation of 10.6\% is noticeable. 
The 3-class approach improves OSD results in all single-channel datasets, particularly on ALLIES (+5.3\%).
Results on ALLIES should be treated cautiously as the average proportion of overlap is rather low, and we identified some issues in the manual segmentation.
In summary, except for ALLIES, the joint VAD+OSD system offers better performance than the two dedicated systems.
It even outperforms the previous state-of-the-art results on DIHARD and AMI data with a new F1-score at 66.8\% and 80.4\% respectively.


\subsection{Domain adaptation}

\vspace{-5pt}
\begin{table}[htbp]
\centering
\caption{VAD and OSD F1-score~(\%) obtained on the ALLIES evaluation set. The model trained on DIHARD is fine-tuned on the subset ALLIES-clean.}
\begin{tabular}{@{}lccc@{}}
\toprule
Model                        & Task & DIHARD         & ALLIES        \\\midrule
\textit{Fine-tuning}                 &      & \textit{ALLIES-clean}       & \textit{No} \\                            
\midrule
\multirow{2}{*}{2-class }   & VAD  & 99.7           & \textbf{99.8} \\
                            & OSD  & 75.3           & 71.6          \\
                             \midrule
\multirow{2}{*}{3-class}    & VAD  & \textbf{99.8}  & 89.2          \\
                            & OSD  & 75.0           & \textbf{75.4} \\
                              \bottomrule
\end{tabular}

\label{tab:adapt}

\end{table}
The presence of errors in the reference segmentation of ALLIES introduces some noise during the training stage, and thus, degrades the performance of the 3-class approach especially regarding VAD.
To cope with this issue, we propose to use the model trained on DIHARD and fine-tune it with the clean subset ALLIES-clean.
Table~\ref{tab:adapt} shows that fine-tuning on ALLIES-clean brings a similar OSD performance (75.0\%) as a model trained with ALLIES data only (75.4\%). 
More interestingly, fine-tuning significantly improves the VAD performance with a relative +11.9\% gain on the F1-score, with only 6~h of in-domain speech.
This gain can be explained by the diversity and quality of the annotations in DIHARD.
We conclude that it is better to train the model on clean and diverse data and apply fine-tuning on in-domain data.


\subsection{Multiple channels}

On the AMI meeting corpus, we notice lower performances on multi-channel data AMI$^\star$ in comparison to the close-talk recordings of AMI. 
Two factors can explain this degradation.
First, multi-channel signals are recorded under distant speech conditions.
This leads to lower quality recordings and thus performance degradation \cite{mariotte22_interspeech}.
Moreover, unlike single channels, the multi-channel feature extraction algorithm does not rely on pre-trained features.
Therefore, the SACC features are less optimized compared to WavLM features.
On AMI$^\star$, the joint VAD+OSD system offers similar VAD performance as the 2-class approach.
The same behavior is observed on the OSD task where the 3-class system degrades with a 0.5\% relative F1-score degradation. 
A single 3-class VAD+OSD system thus offers similar performance as two dedicated VAD and OSD systems on multi-channel audio from AMI$^{\star}$.
VAD and OSD performance are also evaluated on audio data recorded during dinner parties with the CHiME-5 dataset.
Again, the 3-class VAD+OSD system offers similar VAD and OSD performance as two dedicated VAD and OSD systems with about 0.5\% relative F1-score degradation on each task.
Results on these two multi-microphone datasets show that joint VAD+OSD is also adapted to the distant speech scenario with SACC features.

\vspace{-5pt}

\section{Analysis}
\label{sect:5_analysis}

This section evaluates the benefits of such an approach in terms of training time, speech domains, and spatial information in the multi-channel scenario.
\vspace{-5pt}
\subsection{Training time}

\begin{figure}[htp]
    \centering
    \includegraphics[width=\linewidth]{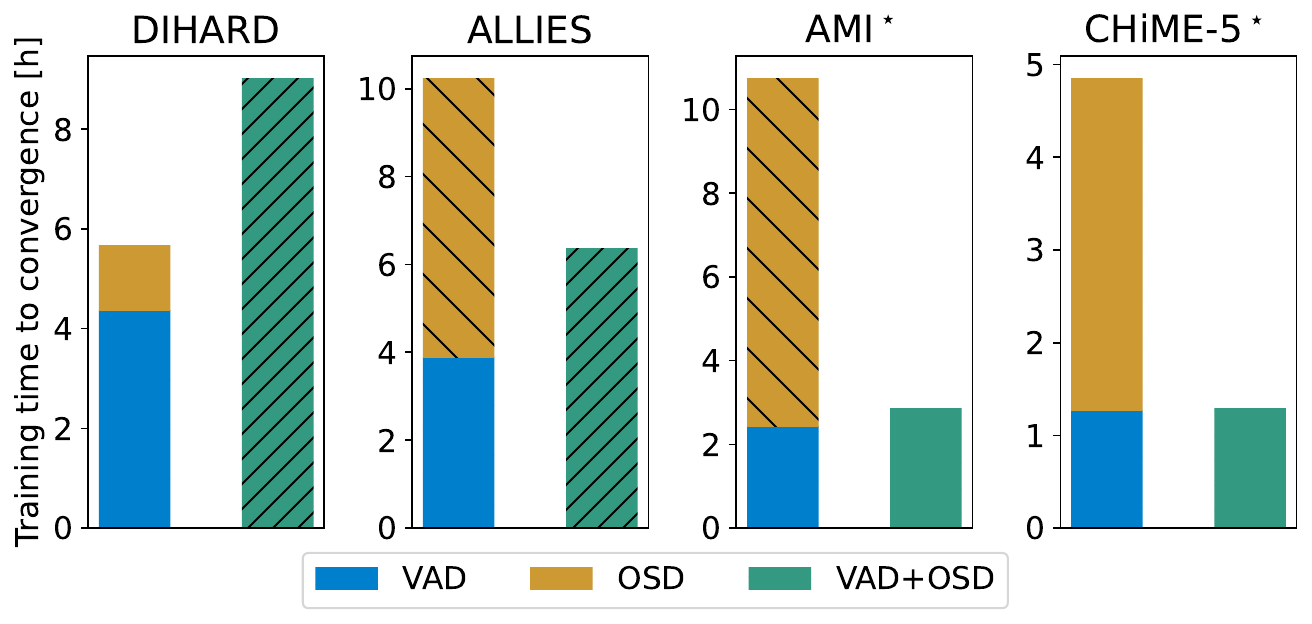}
    \caption{Training time for each system to converge on two single-channel and two multi-channel datasets.}
    \label{fig:train_time_chart}
\end{figure}

In order to assess the value of training a joint VAD+OSD system against two dedicated models, we compare the training time required for each approach.
Each system is trained on an RTX6000 GPU card until it reaches its best F1-score on the validation set.
Figure \ref{fig:train_time_chart} presents the elapsed time to obtain the best-performing model.

2-class OSD task clearly requires more resources than VAD.
Indeed the discrimination of the spectral information between the presence of one speaker or several speakers is more difficult than between speech and non-speech signals.
Multi-channel VAD+OSD system converges as fast as the 2-class VAD system, as observed on the AMI$^\star$ and CHiME-5$^\star$ datasets.
In the single channel scenario, the gain is less significant (and no gain at all with DIHARD), probably because the spatial information helps to detect multiple speakers.

\subsection{Influence of the speech domain on performance}

\begin{figure}[ht!]
    \centering
    \includegraphics[width=\linewidth]{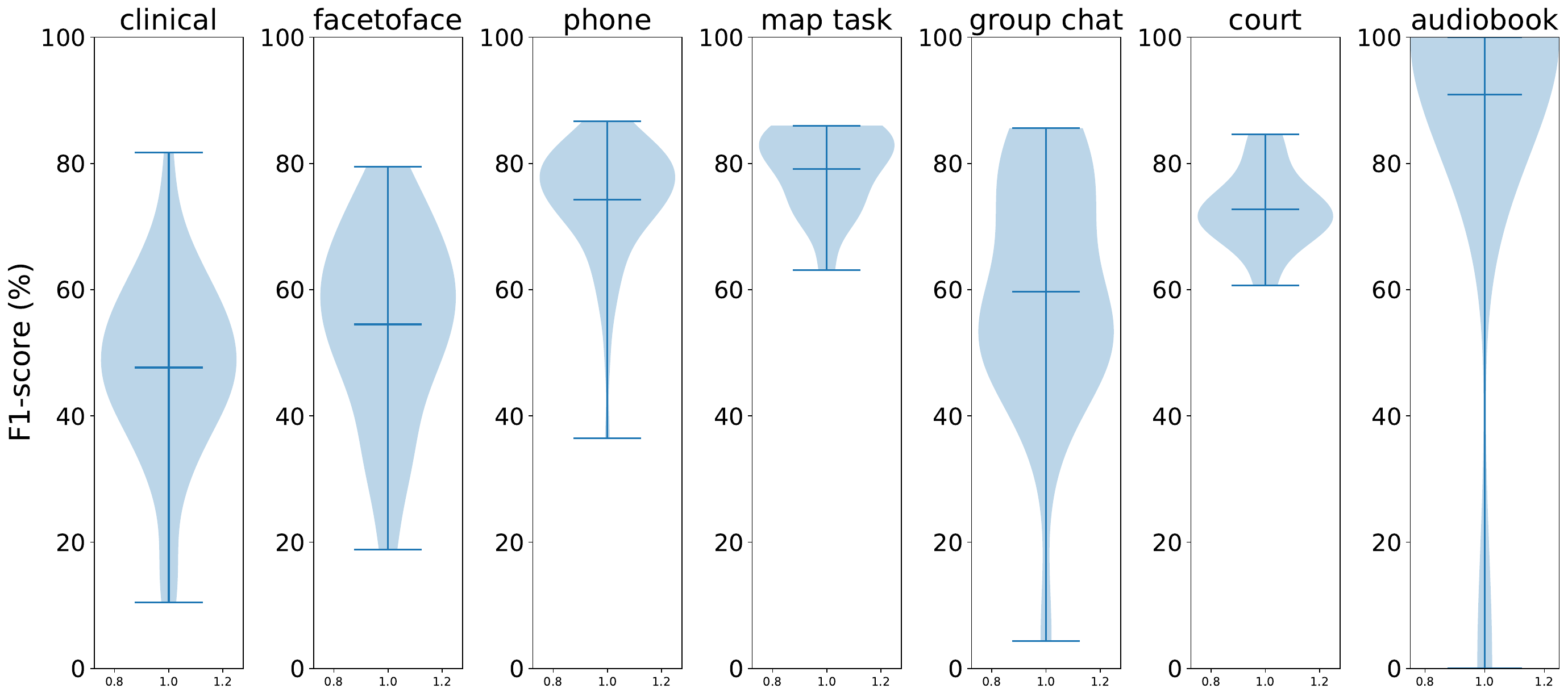}
    \caption{Distribution of F1-scores on DIHARD speech domains}
    \label{fig:domains}
\end{figure}

In order to study the influence of the speech domain on OSD, we analyze the OSD F1-score distributions for each of the DIHARD evaluation files, manually separated into 7 domains (see Fig.~\ref{fig:domains}). \textit{Clinical} contains conversations between a clinician and a child, \textit{facetoface} contain interviews, \textit{phone} contains phone conversations, \textit{map task} contains a game in which someone guides a person remotely on a map, \textit{group chat} contains spontaneous conversations, \textit{court} contains court recordings and \textit{audiobook} contains read speech.

Fig.~\ref{fig:domains} shows that the F1-score is globally better for \textit{phone} conversations than for \textit{clinical} and \textit{face-to-face} conversations, despite the fact that the three domains are dyadic interactions. 
We can then hypothesize that the absence of visual cues in phone conversations limits the diversity of overlaps contained in the audio files. 
Another difference between domains is the quality of the recordings. 
For example, \textit{group chat} and \textit{face-to-face} files 
feature strong background noise and low-quality recordings, which could explain the low performance obtained in these domains.
This analysis concludes that the speech domain is of major importance for OSD.
The presence of noise, the diversity of overlaps, and the differences in turn-taking driven by the speech domain is clearly a major issue for OSD.\\ 

\vspace{-10pt}
\subsection{Spatialisation}

\begin{figure}[ht!]
\centering
\begin{subfigure}{0.32\linewidth}
    \includegraphics[width=\textwidth]{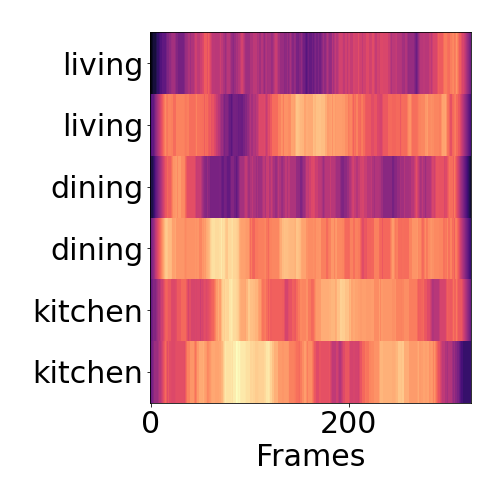}
    \label{fig:first}
\end{subfigure}
\begin{subfigure}{0.32\linewidth}
    \includegraphics[width=\textwidth]{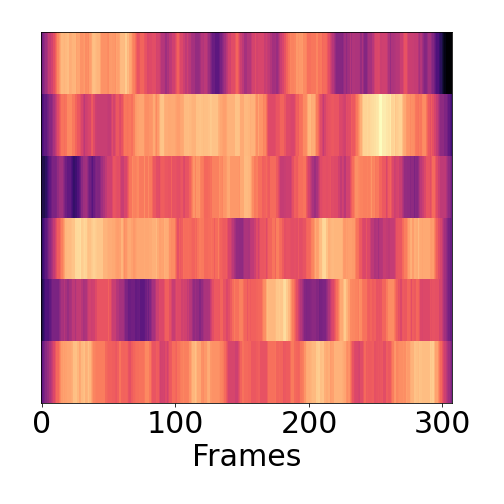}
    \label{fig:second}
\end{subfigure}
\begin{subfigure}{0.32\linewidth}
    \includegraphics[width=\textwidth]{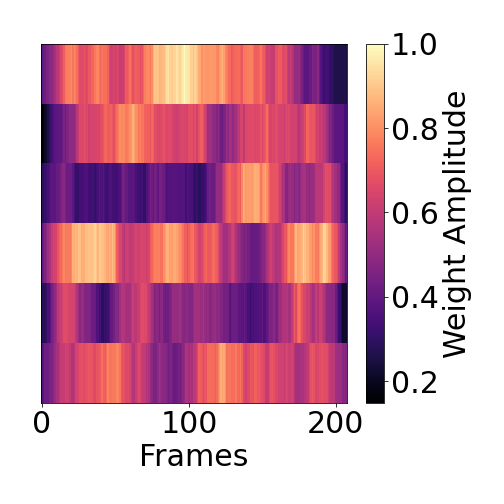}
    \label{fig:third}
\end{subfigure}
\vspace{-15pt}   
\caption{Combination-weights applied to each channel depending on the room where speakers are staying on 3 CHiME-5 utterances. (Left) Kitchen, (Middle) Dining room, (Right) Living room. Utterance-wise normalization is applied for better visualization by dividing weights by their maximum value.}
\label{fig:spat}

\end{figure}

In the CHiME-5 dataset, the rooms where participants are located are annotated for each utterance in the evaluation set.
We can thus study which microphone the SACC feature extractor activates as a function of the speakers' positions.
Since the VAD+OSD system is trained using one microphone per array in the CHiME data, we can visualize the combination weights for each array in each room.
Two arrays are located in the kitchen, two are located in the dining area and two are in the living room.
Figure~\ref{fig:spat} shows the SACC combination weights of each channel depending on the location of the speakers.
On these utterances, the SACC system mostly activates the channels placed in the areas where speakers are located.
The system seems able to select microphones with the most information for the VAD+OSD task.
An in-depth study should however be conducted to better assess the information used by the system in the multiple-channel scenario.

\vspace{-0.5cm}
\section{Conclusion}
\label{sect:6_ccl}
This article presents a benchmark on two speech segmentation tasks -- Voice Activity Detection and Overlapped Speech Detection -- over multi/mono channel and various domains in 5 datasets.
Two approaches are compared by solving jointly or independently VAD and OSD.
The VAD+OSD joint training offers similar performance as the traditional 2-class OSD or VAD approaches on both single-channel audio data and distant multi-microphone signals.
The proposed system reaches a new state-of-the-art for OSD on DIHARD (66.8\%) and AMI (80.4\%) data.
Particularly in the case of ALLIES data with domain adaptation, joint training brings an improvement of +11.8\% for VAD and +5.3\% for OSD.
Furthermore, joint training requires fewer resources as it reduces the training time on most of the datasets, especially in the case of multi-channel data.

A deeper analysis demonstrates that background noise and face-to-face conversations are clearly hard to segment. We also visualize how the combination weights obtained with the SACC multi-channel feature extractor are prone to locate active speakers within a session.

Since VAD and OSD performances on multi-microphone data highly depend on the number of microphones during training, we intend to evaluate our system in a cross-domain scenario with different types of adaptation to go towards a robust multi-corpus segmentation model.
The impact of the proposed VAD+OSD system on diarization will also be evaluated.




\section{Acknowledgments}

This work was performed using HPC resources from GENCI–IDRIS (Grant 2022-AD011012565), the French ANR GEM (ANR-19-CE38-0012), and LMAC grant from Région Pays de la Loire.

\bibliographystyle{IEEEtran}
\bibliography{bib}

\end{document}